\documentclass[aps,twocolumn,groupedaddress]{revtex4}
\pdfoutput=1

\usepackage[pdftex]{graphicx}
\DeclareGraphicsExtensions{.pdf, .jpg}

\begin{document}
% You should use BibTeX and apsrev.bst for references
\bibliographystyle{apsrev}

%\preprint 

\title{Calculations of the Exciton Coupling Elements Between the DNA Bases Using the Transition Density Cube Method}
\author{Arkadiusz Czader}
\affiliation{Department of Chemistry, University of Houston, Houston TX 77204}
\author{Eric R. Bittner}
\thanks{John S. Guggenheim Fellow (2007)}
\affiliation{Department of Chemistry, University of Houston, Houston TX 77204}
\date{\today}

\begin{abstract}
Excited states of the of the double-stranded DNA model (A)$_{12}\cdot$(T)$_{12}$
were calculated in the framework of the exciton theory. The off-diagonal elements 
of the exciton matrix were calculated using the transition densities and ideal 
dipole approximation associated with the lowest energy $\pi\pi^{*}$ excitations 
of the individual nucleobases obtained from TDDFT calculations. The values of 
the coupling calculated with the transition density cubes (TDC) and ideal-dipole 
approximation (IDA) methods were found significantly different for the small 
inter-chromophore distances. It was shown that the IDA overestimates the 
coupling significantly. The effects of the structural fluctuations were 
incorporated by averaging the properties of the excited states over 
a large number of conformations obtained from the MD simulations.
\end{abstract}

%\pacs 
%\maketitle must follow title, authors, abstract and \pacs
\maketitle

\section{Introduction}

DNA a remarkable carrier of code of life, is very stable with respect to
the photochemical decay. The path chosen by Nature to protect DNA is
through the very rapid decay pathways of the electronic excitation
energy. Given the importance of DNA in biological systems and its emerging role
as a scaffold and conduit for electronic transport in molecular
electronic devices, \cite{Kelley:1999} DNA in its many forms is a well
studied and well characterized system.  What remains poorly
understood, however, is the role that base-pairing and base-stacking plays in
the transport and migration of the initial excitation along the double
helix.\cite{Crespo-Hernandez:2005,Markovitsi:17130,Markovitsi:2006}

The absorption of UV radiation by DNA initiate a number of 
 photochemical reactions that can ultimately
lead to carcinogenic mutations.
\cite{Besaratinia:2005,Sutherland:1976,Callis1979qe,Sinha:2002,Freeman:1989}
The UV absorption spectrum of DNA largely
represents the weighted sum of the absorption spectra of it
constituent bases. However, the distribution of the primary
photochemical products of UV radiation, including bipyrimidine dimers,
\cite{Mouret:2006} is depends quite strongly upon  base sequence, 
which implies some degree of coupling between the DNA bases.
\cite{Markovitsi:17130} Inasmuch as both the
base stacking and base pairing are suspected to mediate the excess of
electronic excitation energy, understanding of the excited-state
dynamics is of primary importance for determining how the local
environment affects the formation of DNA photolesions.

Recent work by various groups has underscored the different roles 
that base-stacking and base-pairing play in mediating the fate of 
an electronic excitation in DNA. \cite{Markovitsi:17130,Crespo-Hernandez:2005} 
Over 40 years ago, L{\"o}wdin discussed proton tunneling between bases as 
a excited state deactivation mechanism in DNA\cite{Lowdin:1963} and evidence 
of this was recently reported by Schultz {\em et al.} \cite{Schultz:2004} 
In contrast, ultrafast fluorescence of double helix 
poly(dA){\textperiodcentered}poly(dT) oligomers by Crespo-Hernandez 
{\em et al.}\cite{Crespo-Hernandez:2005} and by Markovitsi {\em et al.} 
\cite{Markovitsi:17130} give compelling evidence that base-stacking rather 
than base-pairing largely determines the fate of an excited state in DNA 
chains composed of adenosine and thymine bases with long-lived 
intrastrand states forming when ever adenosine is stacked 
with itself or with thymine. However, there is considerable debate
regarding whether or not the dynamics can be explained via purely
Frenkel exciton models ~\cite{Emanuele:2005a,Emanuele:2005,Markovitsi:2006} 
or whether charge-transfer states play an intermediate role. 
\cite{Crespo-Hernandez:2006}

Upon UV excitation, the  majority of excited molecules shows 
a subpicosecund singlet lifetimes.
\cite{Pecourt,Pecourt:2000,Gustavsson:2002,Peon:2001}
Owing to the technical difficulties in measuring the ultrashort lifetimes
the study of the charge and excitation energy
transfer in DNA has only recently received much of attention with the
advances in the femtosecond spectroscopy. Although, so far, no clear
picture of the excited -state deactivation mechanism has been offered
by the experiment, two possible decay channels have been investigated.
Kohler and coworkers in their recent study of the duplex
poly(dA){\textperiodcentered}poly(dT) suggested that ${\pi}$-stacking
of the DNA base determines the fate of a singlet electronic excited
state.\cite{Crespo-Hernandez:2005} Alternative decay mechanism involves 
interstrand hydrogen or proton transfer. Douhal and coworkers observed 
excited-state proton transfer in base pair mimincs in gas-phase.
\cite{Douhal:1995} The experimental results suggests that these very 
fast decay pathways play an important role in quenching the reactive 
decay channels and providing DNA with intrinsic photochemical 
stability. However, they do not provide a clear picture which 
arrangement of bases, pairing or stacking, is of primary importance. 

Until recently, most theoretical investigations of excitation energy 
transfer in DNA helices has been within the Frenkel exciton 
model which treats the excitation as a coherent hopping 
process between adjacent bases.\cite{Shapiro:1975,Suhai:1984}
This model has tremendous appeal since it allows one to 
construct the global excited states (i.e of the complete chain) 
in terms of linear combinations of local excited states. 
The key parameter in the evaluation of the electronic excitation energy
transfer (EET) is the electronic coupling between the individual bases.
To a first-order approximation, the base to base coupling 
 can be estimated using a dipole-dipole approximation
 in which the interaction between the donor and acceptor
is calculated using only the transition dipole associated with each
chromophore. While this approach is certainly suitable for cases in which the 
distance between the donor and acceptor sites is substantially greater than the 
molecular length scale. In case of double stranded DNA, where the DNA
bases are in relatively close contact compared to their dimensions this
approach leads to the neglect of the effect of the size and spatial
extent of the interacting transition densities associated with each
chromophore.   

By far the most precise way to calculate the coupling elements is to 
directly integrate the Coulomb coupling matrix element between transition 
densities localized on the respective basis.\cite{Krueger:1998} The 
accuracy is then limited only by the numerical quadrature in 
integrating the matrix element and by the level and accuracy
of the quantum chemical approach used to construct the transition 
densities in the first place. Futhermore, one must perform a  
quantum chemical evaluation of the coupling elements between 
each base at each snapshot along a molecular dynamics simulation 
in order to properly take into account the fluctuations and 
gyrations of the chain itself.  This is a formidable task, one 
that has prevented an accurate benchmarking of the excited state 
electronic structure of realistic DNA chains.  

In this paper, we present the results of simulations and calculations of 
accurate interbase exciton couplings for A-T strands of DNA in water 
in an attempt to provide such a benchmark. Starting from a molecular 
dynamics simulation of a model DNA sequence in water at the correct 
salt concentrations, we mapped out the evolution of the photochemically 
relevant excited states within a Frenkel exciton model in which the 
couplings were computed using both the ideal dipole-dipole approximation 
(IDA) and using the transition density cube approach (TDC).\cite{Krueger:1998}

\section{Methodology}
The calculation procedure consisted of several steps.  In the first stage
the molecular dynamics (MD) calculations were carried out to sample a
range of conformations of (A)$_{12}\cdot$(T)$_{12}$
model of DNA double-helix. The transition densities of the individual
nucleobases obtained from time dependent density functional theory
(TDDFT) calculations were subsequently superimposed with the
instantaneous conformations from the MD simulations in order to
calculate the coupling between the electronic transitions of the
individual bases. In the final step,  the excited-states of the model
were calculated within the Frankel exciton model.

\subsection{Exciton model}
The excited states of the $(A)_{12}\cdot(T)_{12}$ were calculated in 
the framework of the exciton theory\cite{Frenkel:1931, Davydow:1962}. 
In this approach the total Hamiltonian for the super system of 
$N$ molecules is written as the sum of $N$  Hamiltonians of isolated 
molecules $H_{n}$ and the intermolecular interaction potential 
$V_{nm}$ between the molecules $n$ and $m$.
\begin{eqnarray}
H=\sum_{n=1}^{N}H_{n}+\sum _{n=1}^{N}\sum _{m>n}^{N}V_{\mathit{nm}}
\end{eqnarray}
The singly excited states of the system are described in term of
$N$ locally excited configurations
\begin{eqnarray}
\Phi _{l}^{i}=\phi _{l}^{i}\prod _{n\neq l}\phi _{n}
\end{eqnarray}
where $\phi_{l}^{i}$ corresponds to the excited state wavefunction 
of the chromophore $l$ whereas all the other molecules $m$ are in 
their ground state $\phi_n$. $\Phi _{l}^{i}$ denotes the 
corresponding wave function of the super system. Consquently, 
the exciton states of the supramolecular system can be written 
as a linear combinations of the excited states localized on each 
chromophore.
\begin{eqnarray}
\Phi _{k}=\sum_{l}c_{\mathit{kl}}\left|\Phi _{l}^{i}\right\rangle
\end{eqnarray}

The {\em diagonal elements} of the exciton matrix
$\langle \Phi _{n}^{i}|H|\Phi _{n}^{i}\rangle $ are simply 
excitation energies of chromophore $n$ from its ground to
i$^{th}$ excited state, $S_{0}\rightarrow S_{i}$. 
The {\em off-diagonal} elements  
$\langle\Phi_{n}^{i}|H|\Phi_{m}^{j}\rangle $ written as 
$\langle\phi_{n}^{i}\phi_{m}^{0}|V|\phi_{n}^{0}\phi_{m}^{j}\rangle$
correspond to exciton coupling. It can be interpreted as the
electrostatic interaction energy between the transition densities
corresponding to $S_{0}\rightarrow S_{i}$ and $S_{0} \rightarrow S_{j}$.

A measure of delocalization of the exciton states can be obtained from
the inverse participation ratio (IPR)	($1/L_{k}$) which represents the
number of coherently coupled chromophores in a given eigenstate $k$. 
In the general case with more than one electronic transition per 
chromophore, $L_{k}$ is written as follows:
\begin{eqnarray}
L_{k}=\sum_{molecules\ m} \left[ \sum_{states\ i} \left( C_{\mathit{k,m}}^{\mathit{i}} \right)^{2} \right]^{2}
\label{ipr}
\end{eqnarray}
where $k$ denotes a given eigenstate and $i$ an electronic excited state
of a chromophore.

\begin{figure}[t]
\includegraphics[width=\columnwidth]{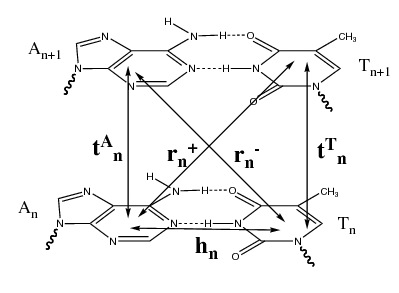}
\caption{Schematic view of the exciton coupling terms for Eq.~\ref{su2}.}
\label{fig3}
\end{figure}

For purposes of developing a model, we can cast the exciton 
Hamiltonian as a $SU(2)\otimes SU(2)$ lattice model \cite{bittner:094909}
consisting of localized hopping interactions for exctions
between adjacent base pairs along each strand ($t_{aj}$) as well as
cross-strand terms linking paired bases ($h_i$) and ``diagonal'' terms
which account for the $\pi$ stacking interaction between base $j$ on
one chain and base $j\pm 1$ on the other chain ($r^\pm_i$) in which
$r^-_j$ denotes coupling in the 5'-5' direction and $r^+_j$ coupling
in the 3'-3' direction.  Fig. ~\ref{fig3} shows a schematic view of
the various coupling terms between each nucleotide base.  
\begin{eqnarray}
H = \sum_j \epsilon_j \hat{\psi}_j^\dagger\hat\psi_j
+ t_j (\hat\psi_{j+1}^\dagger\hat\psi_j + \hat\psi_{j}^\dagger\hat\psi_{j+1}) )
+ h_j\overline{\psi}_j\hat\psi_j \nonumber \\
+\hat\psi_{j+1}^\dagger(r_j^+\hat\gamma_+ + r_j^-\hat\gamma_-)\hat\psi_j
+\hat\psi_{j}^\dagger(r_j^+\hat\gamma_+ + r_j^-\hat\gamma_-)\hat\psi_{j+1},
\label{su2}
\end{eqnarray}
where $\hat\psi_j^\dagger$ and $\hat\psi_j$ are $SU(2)$ spinors that
act on the ground-state to create and annihilate excitations on
the $j$th adenosine or thymidine base along the chain.  The
$\hat\gamma$ operators are the $2\times 2$ Pauli spin matrices with
$\overline{\psi}_j = \hat\gamma_1\hat\psi_j^\dagger$ and
$\hat\gamma_++\hat\gamma_- = \hat\gamma_1$ providing the mixing
between the two chains.  

Taking the chain to homogeneous and infinite
in extent, one can easily determine the energy spectrum 
of the valence and conduction bands by diagonalizing
\begin{eqnarray}
\hat{H}(q)=
\left(
\begin{array}{cc}
\epsilon_{A} + 2 t_{A} \cos(q)   &  h + r^{+}e^{-iq} + r^{-}e^{+iq} \\
h + r^{+}e^{+iq} + r^{-}e^{-iq} \    & \epsilon_{T} + 2t_{T} \cos(q)
\end{array}
\right)
\label{bandop}
\end{eqnarray}
where $\epsilon_{A,B}$ and $t_{A,T}$ are local excitation 
energies and intra-strand hopping integrals.  $h$ is the 
coupling between Watson-Crick bases. When the interchain 
diagonal couplings are equal, $r^+ = r^-$, Eq.~\ref{bandop} 
is identical to the Hamiltonian used by Creutz and Horvath
~\cite{Creutz:1994} to describe chiral symmetry in quantum 
chromodynamics in which  the terms proportional to $r$ are 
introduced to make the ``doublers'' at $q\propto \pi$ heavier 
than the states at $q \propto 0$ since the off-diagonal coupling 
is now momentum dependent. 

One of the {\em serious} deficiencies with this model as it stands thus far is that
for DNA each of the interactions described is very sensitive to the 
geometric fluctuations of the DNA chain itself. \cite{Emanuele:2005,Emanuele:2005a}
Hence, we need to consider each of the couplings as being parametrically dependent
upon the instantaneous molecular geometry of both the individual bases and the 
chain itself.  This is assuming there is no additional contribution from the solvent and
ions surrounding the DNA chain.    Assuming that the electronic time scale is fast 
compared to the typical time scale for geometric fluctuations of the DNA chain 
($10^{-14} - 10^{-13}$s for longitudinal and 
$10^{-13} - 10^{-12}$ s for the lateral motions  of bases in DNA 
double helices\cite{McCammon:1987}), we can consider at least the initial 
electronic dynamics as occurring in a fixed nuclear framework and subsequent 
dynamics as adiabatically following the nuclear motion.  
Nonadiabatic contributions can not be completely discounted; however, 
the dominant non-adiabatic couplings are intermolecular in origin or involve
proton between adjacent bases. \cite{Clelia:2005,Sobolewski:2002,Lowdin:1963}

\subsection{Transition densities and interactions}

\subsubsection{Exciton-exciton interactions}
Each off-diagonal term in our Hamiltonian of Eq.~\ref{su2} can 
be calculated according to
\begin{eqnarray}
V^{\mathit{Coul}}=\sum _{ab}{\frac{M_{n}^{0i}(a)M_{m}^{0j}(b)}{4\pi
\epsilon _{0}r_{ab}}}
\label{coul}
\end{eqnarray}
where the two terms in the numerator, $M_{n}^{0i}$ and  $M_{m}^{0j}$ 
are the three dimensional charge distributions (transition densities) 
associated with the ground and electronic excited states $i$ and
$j$ of molecules $n$ and $m$, respectively, with the separation 
between the elements $a$ and $b$ equal to $r_{ab}$. 
The $V^{\mathit{coul}}$ corresponds to the electrostatic repulsion 
energy between the two charge distributions $M_{n}^{0i}$ and $M_{m}^{0j}$
of isolated chromophores. The calculations of the Coulombic couplings 
using the three dimensional charge distribution takes into account 
the size and the spatial extent of the transition density and
is valid at all molecular separations as opposed to the {\em ideal
dipole approximation} (IDA). In the latter only the dipole moment of
the transition density is considered for calculations of the coupling
terms which makes the computations of the off-diagonal elements much
more efficient. However, this approximation breaks down at the small
donor-acceptor separations for which the spatial extent of the transition
density becomes important.

 \begin{figure}[b]
 \includegraphics[width=\columnwidth]{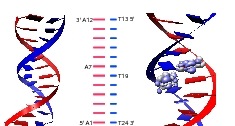}
 \caption{Schematic structure of the (A)$_{12}\cdot$(T)$_{12}$ 
 oligomer used in the MD simulations (left). The residue numbering 
 is shown in the middle. The graphic on the right shows the 
 three dimensional representations of densities corresponding 
 to lowest energy ${\pi\pi^*}$ transitions of adenine and thymine 
 superimposed with residues 7 and 19, respectively, of the 
 (A)$_{12}\cdot$(T)$_{12}$ model.}
 \label{fig_trans_dens_overlapped}
 \end{figure}

To account for the dynamics of the DNA chain itself, we performed a 
series of molecular simulations of the 12 base pair duplex DNA (AT) 
(Figure \ref{fig_trans_dens_overlapped}) with about 12,000 water 
molecules and counter ions. 
\footnote{
Regarding the molecular dynamics simulations: the simulation was performed 
with the extended system (ESP) molecular dynamics program \cite{ESP}. 
The system consisted of a 12 base pair duplex DNA (AT) with $11,593$ 
waters, $46$ sodium ions, and $24$ chloride ions in a cubic box of 
length $70.4$ \AA. The atomic interactions were defined by the 
CHARMM (version $27$) force field \cite{Charmm}. The system was minimized and 
equilibrated in the NVE ensemble at $300$K. The bonds were kept rigid using 
the Rattle \cite{Rattle} implementation of the Shake method \cite{Shake} and 
the electrostatic interactions were evaluated using the Ewald sum technique
\cite{Ewald}. The equations of motion were integrated using the Velocity
Verlet algorithm \cite{Vverlet} with a $2$ femtosecond time step. The 
simulation was initially run for $15$ nanoseconds. Next, the timestep was 
changed to $1$ femtosecond and the snapshots saved every $10$ steps.
}
Once the system was minimized and equilibrated in the NVE ensemble 
at $300$K, we integrated the dynamics for an additional 80 ps, 
sampling the DNA configuration every 10 fs. Even though we are dealing 
with a relatively small strand, it remains too large for an accurate 
evaluation of its electronic structure. Consequently, we make the 
approximation that the excited states of the molecule itself can 
be written as a linear combination of excited states localized 
on the instantaneous positions of each base along the chain. 
Furthermore, given the computational cost associated with evaluating 
the excited states of even a small molecule, it is prohibitive to 
perform such calculations for each base at each time-step. Our approach,
then, is to perform an accurate evaluation of the local transition 
densities based upon the geometries of the isolated DNA bases, then 
map these densities onto the instantaneous positions of the bases from 
the molecular dynamics simulations (Figure \ref{fig_trans_dens_overlapped}). 
From this, we can evaluate the exciton-exciton coupling (Eq.~\ref{coul}) 
in which $M_{n}^{0i}$ and  $M_{m}^{0j}$  are the transition densities 
about the instantaneous positions of bases $n$ and $m$. 

\begin{figure}[t]
\includegraphics[width=\columnwidth]{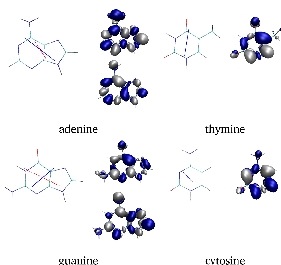}
\caption{The transition densities and transition moments in the 
  nucleobases: adenine, thymine, guanine, and cytosine. For purines 
  the solid and dotted lines indicate the direction of the transition 
  dipoles associated with the first and second lowest energy $\pi\pi^*$ 
  transitions, respectively.}
\label{fig1}
\end{figure}

\subsubsection{Excited states of individual bases}
The geometries of the DNA bases, adenine, guanine, cytosine, 
and thymine in their most common tautomeric forms were optimized 
at the MP2/TZVP level of theory in chloroform using Gaussian03 
suite of programs.\cite{g03} The optimized geometries were 
subsequently used to calculate the singlet excitation energies in 
gas phase at the TD-DFT level using PBE0 functional and TZVP basis 
set augmented with the diffusion functions on all atoms as implemented 
in ORCA.\cite{ORCA} Additionally, the excitation energies were also 
calculated for the standard nucleobase geometries obtained from the 
X3DNA.\cite{x3DNA} In these calculations the deoxyribose and 
phosphate groups were replaced with hydrogens using the 
Chimera program.\cite{Chimera} Without further optimization of 
the structures, the excitation energies were calculated at the same 
level of theory as used before for the MP2 optimized structures. 
Fig.~\ref{fig1} shows both the transition density  and direction 
of the transition dipole moment for each base as given by TDDFT 
after optimization at the MP2/TZVP level in a CHCl$_3$. Transition 
moments were calculated using TDDFT with PBE0 functional and aug-TZVP
basis set in vacuum. The calculated excitation energies are summarized 
in Table~\ref{table1}.

\begin{table*}[t]
\caption{Vertical singlet excitation energies (eV) followed by oscilator
  strength in parentheses of the lowest electronic transitions of the nucleobases 
  calculated using TDDFT at PBE0/aug-TZVP level of theory and MRCI in vacuum for the 
  standard geometries taken from 3DNA and optimized at MP2/TZVP level in chloroform. 
  For the calculated excited state energies the transitions with a $\pi\pi^*$ character
  are indicated with a boldface.}
\label{table1}
\begin{ruledtabular}
\begin{tabular}{lllccccc}
  & Method     & Geometry & $S_{0}\rightarrow S_{1}$ & $S_{0}\rightarrow S_{2}$ & $S_{0}\rightarrow S_{3}$ & $S_{0}\rightarrow S_{4}$ & $S_{0}\rightarrow S_{5}$ \\\hline
A & TDDFT      & MP2      &      5.00 (0.002)  & {\bf 5.29 (0.230)} & {\bf 5.38 (0.069)} &      5.49 (0.009)  &       \\ 
  &            & standard &      5.17 (0.001)  & {\bf 5.44 (0.204)} &      5.46 (0.006)  & {\bf 5.52 (0.086)} &       \\
  & MRCI       & MP2      & {\bf 4.80 (0.168)} &      5.01 (0.003)  & {\bf 5.13 (0.446)} &      5.13 (0.004)  &       \\
  & Exp.$^{a}$ &          &      4.5{--}4.6    &      4.7{--}4.9    &      5.8{--}6.1    &                    &       \\
\hline
G & TDDFT      & MP2      &      4.84 (0.024)  & {\bf 5.07 (0.142)} &      5.22 (0.002)  & 5.24 (0.014)       & {\bf 5.42 (0.304)}\\
  &            & standard &      4.58 (0.001)  & {\bf 5.04 (0.167)} &      5.08 (0.002)  & 5.25 (0.000)       & {\bf 5.41 (0.313)}\\
  & MRCI       & MP2      &      3.68 (0.001)  & {\bf 4.74 (0.286)} &      5.18 (0.002)  & {\bf 5.21 (0.478)} &      5.57 (0.014) \\
  & Exp.$^{b}$ &          &      4.4{--}4.6    &      4.9{--}5.1    &      5.5           &      6.1{--}6.3    &             \\
\hline
T &  TDDFT     & MP2      & 4.74 (0.000)       & {\bf 5.22 (0.161)} &      5.66 (0.000)  &        &  \\
  &            & standard & 4.74 (0.000)       & {\bf 5.21 (0.156)} &      5.63 (0.000)  &        &  \\
  & MRCI       & MP2      & 4.63 (0.000)       & {\bf 5.35 (0.434)} &      5.74 (0.004)  &        &  \\
  & Exp.$^{c}$ &          & 4.6{--}4.7         &      5.6{--}6.1    &      6.4           &        &  \\
\hline
C & TDDFT      & MP2      & {\bf 4.78 (0.049)} &      4.84 (0.000)  &      5.18 (0.002)  & & \\
  &            & standard & {\bf 4.70 (0.040)} &      4.77 (0.000)  &      5.07 (0.002)  & & \\
  & MRCI       & MP2      & {\bf 4.69 (0.151)} &      4.73 (0.002)  &      5.69 (0.007)  & & \\
  & Exp.$^{d}$ &          &      4.5{--}4.6    &      5.0{--}5.4    &      5.6{--}6.1    & & \\
 \end{tabular}
 \end{ruledtabular}
Average experimental excitation energies from Refs. 
$^{a}$ \cite{Clark:1965,Clark:1990,Voelter:1968}
$^{b}$ \cite{Clark:1994,Voet:1963,Clark:1977,Yamada:1968}
$^{c}$ \cite{Voelter:1968,Yamada:1968,Sprecher:1977,Brunner:1975}
$^{d}$ \cite{Voet:1963,Sprecher:1977,Zaloudek:1985,Raksanyi:1978,Miles:1969}
\end{table*}

The transition densities associated with the allowed ${\pi\pi^*}$
excitations of the individual nucleobases, defined by
\begin{eqnarray}
M_{eg}({\bf r};s) = |\Psi_{g}\rangle \langle\Psi_{e}| dr
\end{eqnarray}
where $e$ and $g$ correspond to the
excited and ground states of the chromophore, were calculated using
ORCA program. The densities are written in form of a charge
distribution over three-dimensional grid of points, such that the
integrated charge vanishes, according to
\begin{eqnarray}
 M_{eg}\left(x,y,z\right)=V_{\delta }\int _{z}^{z+\delta_{z}}
 \int _{y}^{y+\delta _{y}}\int _{x}^{x+\delta _{x}}\Psi _{g}\Psi_{e}^{*}
\text{d}s\;\text{d}x\;\text{d}y\;\text{d}z
\end{eqnarray}
where $V_{\delta} = \delta_{x}\delta_{y}\delta_{z}$
is the element volume and the $\delta_{x}$, $\delta_{y}$, $\delta_{z}$ 
are the steps along the coordinate axes. The grid size has to be 
a compromise between the accuracy and the speed. Denser grids 
render the calculations very taxing while too small grids introduce 
large errors in the calculated coupling elements. A satisfactory 
compromise was obtained for the cube files with 40 voxels along each 
axis ($x$, $y$, and $z$) which corresponds to total number of 
64000 elements. In case of the single nucleobase the volume of a single 
element is than 0.03 {\AA$^{3}$}. The changes in the magnitude of the 
coupling calculated for cubes with number of elements larger than 64000 
was below 0.1 cm$^{-1}$ . Owing to the finite size of the cube the integrated 
charge over space was not exactly zero. The residual charge 
for all the transition density cubes was below 0.01$e$ and was 
compensated by adding equal amount of charge to each volume element 
to bring the integrated charge over the cube volume to zero.

The transition densities between the ground and excited states of the
individual DNA bases were generated using TDDFT at the geometries of
the bases optimized at MP2 level of theory, as described above. Before
the actual calculation of the coupling elements could be carried out
the transition densities and dipole moments obtained from {\em ab
initio} calculations in an arbitrary coordinate system were
transformed to the geometries of the bases in the studied DNA
structures. This was carried out by defining the transformation
superposing the plane defined by C6, N1, and N3 atoms of pyrimidine 
or C6, N1, and N9 atoms of purine bases in the arbitrary system 
with the plane defined by the corresponding three atoms of the 
base in the DNA structure. Subsequently, the transformation was 
applied to the three-dimensional grid holding the transition 
density and the dipole moments. The quality of the fit as
measured by the root-mean squared deviation between the atom 
coordinates of the two overlapped structures was very good.

\section{Results and Discussion}
\subsection{Individual Nucleobases}
The optimization of the standard nucleobase geometries \cite{Clowney:1996} 
obtained from X3DNA \cite{x3DNA} at the MP2 level has very small 
effect on their geometries. The only noticeable difference is 
the out-of-planarity of the NH$_{2}$ groups of adenine, guanine, 
and cytosine in the optimized geometries. This is in agreement 
with previous theoretical studies of Shukla {$et$} {$al.$} 
(\cite{Leszcz:2004} and references therein) and experiment 
\cite{Dong:2002} where the amino groups of Ade, Gua, and Cyt 
were also found to be non-planar.. The root mean square deviations 
(RMSD) between the heavy atoms (excluding hydrogens) of the original 
and optimized structures is 0.045 {\AA} for adenine (Ade), 0.040 
for guanine (Gua), 0.017 {\AA} for thymine (Thy), and 0.024 {\AA} 
for cytosine (Cyt).

\subsubsection{Excited-State Calculations}
The MP2 optimized structures of the 9H-purines and 1H-pyrimidines were subsequently used
to calculate the vertical excitation energies using time dependent density functional theory
(TDDFT) at the PBE0/augTZVP level in gas phase. The results of the excited state calculations
on Ade, Gua, Cyt, and Thy were also compared with the available experimental data and 
multireference configuration interaction (MRCI) calculations. 

{\em Adenine}. The lowest TDDFT calculated vertical singlet excitation energies of adenine,
5.00, 5.29, and 5.38 eV \ (Table~\ref{table1}) correspond to the $n\pi^{*}$ and two closely spaced  
$\pi\pi^{*}$ transitions, respectively. While this order is in agreement with other DFT 
calculations (\cite{Leszcz:2004}), {\em ab initio} calculations at CASPT2 level 
(\cite{Fulscher:1997}) has shown the lowest excited state, $S_{1}$, to be a $\pi\pi^{*}$ 
state in agreement with the experiment. The UV-spectra of 9-methyladenine in stretched 
polymer poly(vinyl alcohol) films collected by Holmen (\cite{Holmen:1997}) show two 
in-plane polarized transitions located at 4.55 and 4.81 eV. Contrary to the TDDFT 
results the experimental data show the low-energy $\pi\pi^{*}$ transition to carry 
less oscillator strength. The higher level {\em ab initio} calculations performed at 
MRCI level also predict the lowest energy state to be the light absorbing $\pi\pi^{*}$  
state, calculated at lower energy, 4.81 eV, compared with TDDFT results. The second $\pi\pi^{*}$ 
state is calculated at 5.07 eV. In accordance with the experiment and CASPT2 data the MRCI 
predicts the higher energy $\pi\pi^{*}$  transition to be more intense than the lower energy one. 
The most noticeable structural change, between the MP2 optimized and standard geometry of Ade is the pyramidalization of the amine N in the former. The calculated transition energies at the 
TDDFT level imposed by the pyramidalization show a blue shift in the range of 0.10{--}0.15 eV 
for the flat structure. However, the separation between the two $\pi\pi^{*}$  states and the 
character of the first excited state are not noticeable changed at this level of theory.

{\em Guanine}. For the structure with the planar geometry of the
NH$_{2}$ group (standard geometry) the computed lowest
excitation energy at 4.59 eV is classified as the $\pi\sigma^{*}$
transition. For this transition the configuration with the highest
percentage weight, 99\%, corresponds to HOMO $\rightarrow$ LUMO, with the
LUMO orbital being a $\sigma^{*}$ localized at NH$_{2}$ group. The
lowest energy $\pi\pi^{*}$ transitions for the flat structure are calculated 
at 5.04 and 5.41 eV. The pyramidalization of the NH$_{2}$ group in the
MP2 optimized structure causes the lowest energy transition to acquires
some ${\pi}$* character. It is now calculated at higher energy 4.89 eV
and defined by the configurations HOMO $\rightarrow$ LUMO (91\%) and HOMO
$\rightarrow$ LUMO+1 (5\%). At this geometry the LUMO orbital is a mixture of
${\pi}$* and ${\sigma}$* and the LUMO+1 is pure ${\pi}$*. Similar
mixing of ${\pi}$* and ${\sigma}$* character was reported by
Leszczynski and coworkers, who assigned corresponding transition for
the nonplanar structure to the weak $\pi\pi^{*}$ transition. In our
calculations the two lowest energy  $\pi\pi^{*}$ transitions are
calculated at 5.07 and 5.42 eV and the transition at 4.89 eV is
classified as  $\pi\sigma^{*}$. For this assignment of the
 $\pi\pi^{*}$ transitions the difference in their calculated energies
for the planar and pyramidal geometry of NH$_2$ group is very small, below
0.05 eV. A completely different situation was observed for the adenine,
for which pyramidalization of the NH$_2$ group caused blue shift of the
$\pi\pi^{*}$ transitions.
The MRCI calculations yield the two lowest singlet vertical excitation energies
at 4.24 and 4.34 eV which have $n\pi$* character. The lowest 
$\pi\pi^{*}$ transitions at this level of theory are calculated at 4.76 and 5.24 eV.
The latter $\pi\pi^{*}$ transition has also larger calculated oscillator strength
similarly to was was observed for adenine at this level of theory. 

{\em Thymine}. The ab initio and TDDFT calculations predict the
lowest energy $n\pi^{*}$ transition in vacuo for Thy in accordance with
reported experimental data. The excitation energies calculated at the
optimized and standard geometries of Thy are virtually the same for both 
the $n\pi^{*}$ and $\pi\pi^{*}$ transitions. 
In an aprotic solvent thymine has the $n\pi^{*}$ state as the lowest 
singlet excited state \cite{Callis:1983}. Present calculations 
at the TDDFT level, show dark $n\pi^{*}$ singlet excited state calculated
at 4.74 eV, approximately 0.5 below bright $\pi\pi^{*}$ state calculated
at 5.22 eV. At the MRCI level the relative order of this two
transitions is the same and the calculated energies 4.63 and 5.35 eV of 
the $n\pi^{*}$ and $\pi\pi^{*}$ transitions, respectively, are in good
agreement with the corresponding TDDFT values.

{\em Cytosine}. The TDDFT computed vertical singlet excitation
energies of cytosine, shown in (Table~\ref{table1}), predict the $\pi\pi^{*}$
state to be the lowest energy transition calculated at 4.78 and 4.70 eV
for MP2 and standard geometries, respectively. The non-planarity of the
NH$_{2}$ has only small effect on the energy of the $\pi\pi^{*}$ transition 
inducing a red shift with a magnitude below 0.1 eV. 
The MRCI calculations predicts the same order of the two lowest transitions, 
with $\pi\pi^{*}$ below $n\pi^{*}$ and their energies within 0.1 eV of the
corresponding TDDFT values (Table~\ref{table1}).

\begin{figure}[t]
\includegraphics[width=\columnwidth]{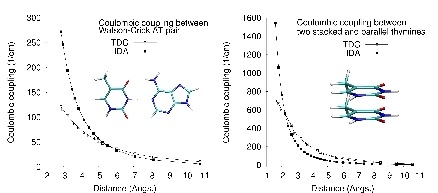}
\caption{Coulombic coupling between the lowest $\pi\pi^*$ transition moments 
  of Watson Crick AT base pair (left) and two stacked and parallel thymines
  (right) as a function of the distance between the two bases. The distance 
  is measured between N1 and N3 atoms of A and T, respectively, for the AT pair
  (left) and between the centers of the mases of the two thymines (right).}
\label{fig2}
\end{figure}
 
\subsection{Coulombic Coupling}
The values of the Coulombic couplings between the lowest energy
$\pi\pi^{*}$ transitions of the adenine and thymine and two
${\pi}$-stacked thymines as a function of distance between the bases
(Fig. \ref{fig2}) were calculated using the TDC and IDA methods. 
For the calculations the transition densities and dipoles were those 
obtained from the TDDFT calculations on the MP2 optimized 
geometry of the basese. The comparison of the coupling 
elements obtained with the two methods, IDA and TDC, (Fig.~\ref{fig2})
shows a good agreement at a separation between the bases larger than
5 and 6 {\AA} for the AT pair and two stacked thyminess, respectively. 
At a shorter
separations, in the range of 3--4 {\AA}, which is typical for DNA
structures, the agreement between IDA and TDC is very poor with the
differences between calculated couplings larger than 100\% in
case of AT pair. The aforementioned good agreement between IDA and TDC
at a larger and poor agreement at shorter separations between
nucleobases indicate that the shape and spatial extent of transition
density (Fig. \ref{fig2}) become important and cannot be neglected at
distances between the bases typical for double helices DNA. The
agreement between the two methods becomes very good in the limit of
very large separation, ($>$ 8 {\AA}).

\begin{table*}[t]
\caption{Average values of the Coulombic couplings 
  (cm\textsuperscript{-1}) between the Adenine and Thymine bases of the
  (A)$_{10}\cdot$(T)$_{10}$ oligomer in the idealized BDNA
  geometry. The reported values are the averages for all
  corresponding base pairs with their standard deviations in parentheses. 
  The $h_{n}$, $t_{n}$, and $r_{n}^{\pm}$ correspond to coupling terms in 
  Eq.\ref{su2} and Eq.\ref{bandop}.}
\label{table}
\begin{ruledtabular}
\begin{tabular}{lllll}
& base 1 & base 2 & IDA & TDC\\
\hline
Watson-Crick Base Pairs\\
$h$ & A & T & 229.5 (0.1) & 100.8 (0.04) \\
\hline
$\pi$-stacked bases: nearest neighbors:\\
$t_{A}$ & A & A & 871.7 (0.4) & 160.7 (0.2)\\
$t_{T}$ & T & T & 201.4 (0.1) & 90.2 (0.03)\\
\hline
$\pi$-stacked bases: 2nd nearest neighbors \\
& A & A & 57.4 (0.1)  & 8.7 (0.02)  \\
& T & T & 3.1 (0.005) & 1.4 (0.002) \\
\hline
interchain diagonal cross terms   \\
$r^{+}$ & A & T & 163.4 (0.03) & 109.1 (0.01) \\
$r^{-}$ & T & A & 90.7 (0.02)  & 25.2 (0.01) \\
\end{tabular}
\end{ruledtabular}
\end{table*}

\begin{figure}[b]
\includegraphics[width=\columnwidth]{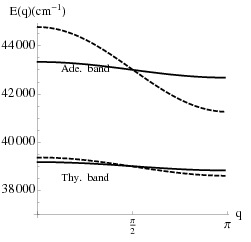}
\caption{Band structure for AT B-DNA.  Dashed curves: IDA, Solid: TDC.}
\label{dnaband}
\end{figure}

In order to compare the performance of the IDA and TDC method we
calculated the coulombic couplings between the lowest $\pi\pi^{*}$ 
transitions of adenine and thymine bases for the interstrand
Watson-Crick, intrastrand ${\pi}$-stacked, and diagonal arrangement 
(Fig.~\ref{fig3}) of the base pairs for the $(A)_{10}\cdot(T)_{10}$ 
oligomer in the idealized B-DNA geometry (Table \ref{table2}) 
generated using the 3DNA program\cite{x3DNA} using the IDA and TDC 
approximations. The corresponding band-structure as given by 
Eq.~\ref{bandop} is shown in Fig.~\ref{dnaband}. 

For the stacking and pairing distances corresponding to
the idealized B-DNA geometry the coupling elements calculated with
the IDA approximation result with several-fold larger absolute values
compared with the corresponding values calculated using the TDC
method. The largest differences between the two methods are
obtained for the couplings between the ${\pi}$-stacked adenines. For
the idealized B-DNA geometry the coupling between two adenines
located on the same strand calculated using IDA, 872 cm$^{-1}$, 
is more than 5-fold larger compared with the value obtained using 
TDC 161 cm$^{-1}$. The differences in the calculated couplings using 
the same two methods for two stacked thymines are much smaller. 
For this base pair the Coulombic coupling calculated using IDA 
is equal to approximately 230 cm$^{-1}$, more than twice the value 
of 101 cm$^{-1}$ obtained with TDC. The absolute values of the 
coupling elements between the second nearest neighbors
located on the same strand are much smaller. At the IDA level of
approximation the coupling between the two adenines is only 57 cm$^{-1}$ 
compared with 9 cm$^{-1}$ calculated for the same base pairs
using transition density cubes. The coupling between the two thymine
bases on the same strand is even smaller {--} approximately 3 and 1
cm$^{-1}$ for IDA and TDC methods, respectively.

The values of the calculated coupling elements give hint on the relative
exciton delocalization along the thymine and adenine strands of the 
$(A)_{10}\cdot(T)_{10}$ oligomer. The band structure in Fig.~\ref{dnaband}
shows that the mobility of the exciton along the thymine strand is low with 
both methods, IDA and TDC, giving resonable close results. The 
exciton mobility along the adenine strand, to the contrary, is quite 
different for the two methods. IDA, in this case, predicts more delocalized 
exciton states compared to TDA, which reflects larger magnitudes of the
couplings calculated with the former method. 

The magnitudes of the couplings between bases located on different
strands (Table~\ref{table2}), which belong to the Watson-Crick base pairs, are
also larger when calculated with IDA method. The average Coulombic
coupling for the Watson-Crick AT base pair calculated using IDA is 230 cm$^{-1}$ 
compared with 101 cm$^{-1}$ obtained with TDC method. The
magnitudes of the couplings between bases located on different strands,
which does not belong to the Watson-Crick, the diagonal terms, are
generally still quite large. Especially, the magnitude of the coupling
between diagonal bases in the 5'-XY-5' direction ($r^+$) is comparable
with the coupling for the Watson-Crick basepairs (Table~\ref{table3}). 
The values computed for the AT pair with the IDA and TDC methods are 163 and 109
cm$^{-1}$, respectively. These values are almost twice and four times larger 
compared with the coupling between the corresponding bases in the 3' direction 
($r^-$)calculated with the IDA and TDC methods, respectively. From the 
results of Table~\ref{table2} it is clear that the coupling elements 
are very sensitive with respect to the base sequence. The calculated 
Coulombic couplings for the ${\pi}$-stacked arrangement of the 
adenine bases are by far the largest. Using IDA method the 
calculated coupling elements are more than two-fold larger than the 
corresponding values calculated with IDA for almost all arrangements.

\begin{table*}[t]
\caption{Average Coulombic coupling $\overline{V}_{ij}$ (in cm$^{-1}$) 
  between selected bases of (A)$_{12}\cdot$(T)$_{12}$. 
  The values are averaged over 4000 snapshots from an 80 ps 
  MD simulation of DNA in water. $\sigma$ is the r.m.s. deviation 
  about the mean.  Max. and Min. refer to the maximum and minimum 
  value of the coupling over the entire simulation.}
\label{table3}
\begin{ruledtabular}
\begin{tabular}{lllrrrr}
&base 1 & base 2& $\overline{V}_{ij}$ & $\sigma$ & Max  & Min \\
\hline
Interchain: Watson-Crick Base pairs  \\
IDA\\
& A9 & T16 & 235.7 & 29.9 & 325.8 & 105.2\\
& A4 & T21 & 237.9 & 31.3 & 347.7 & 100.7\\
& A7 & T18 & 235.8 & 33.3 & 363.0 & 112.0\\
TDC\\
& A9 & T16 & 109.3 & 11.7 & 151.6 & 71.4\\
& A4 & T21 & 101.7 & 12.8 & 146.7 & 51.1\\
& A7 & T18 &  99.7 & 15.3 & 146.3 & 39.3\\
\hline
Intrachain: nearest neighbor\\
IDA \\
& A3 & A4 & 769.3 & 163.5 & 1338.4 & 322.0\\
& A6 & A7 & 858.1 & 183.9 & 1378.9 & 284.2\\
& A9 &A10 & 935.9 & 153.6 & 1396.6 & 434.1\\
&T21 &T22 & 473.3 & 153.7 & 1007.4 &  18.6\\
&T18 &T19 & 435.8 & 136.7 & 1041.8 & 110.9\\
&T15 &T16 & 315.4 & 145.1 &  874.0 & 210.4\\
TDC\\
& A3 & A4 & 175.4 & 47.1 & 330.7 & 50.0\\
& A6 & A7 & 197.2 & 65.4 & 363.7 &  0.0\\
& A9 &A10 & 196.8 & 43.9 & 339.5 & 48.3\\
&T21 &T22 & 164.9 & 45.9 & 338.9 & 38.0\\
&T18 &T19 & 159.8 & 36.5 & 301.8 &  0.0\\
&T15 &T16 & 124.2 & 39.7 & 277.3 & 17.4\\
\hline
Intrachain: 2nd nearest neighbor\\
IDA\\
& A3 & A5 & 62.8 & 14.2 & 104.8 & 12.3\\
&T20 &T22 & 25.1 & 15.5 &  74.2 & 24.8\\
TDC\\
& A3 & A5 & 11.9 & 8.2 & 41.6 &13.8\\
&T20 &T22 &  9.7 & 6.7 & 30.2 &12.2\\
\hline
Interchain; diagonal terms \\
IDA\\
& A8 &T16 & 129.0 &32.3 & 238.0 & 28.2\\
& A9 &T17 & 121.8 &38.8 & 247.0 &  1.9\\
TDC\\
& A8 & T16 & 77.0 & 17.2 & 150.0 & 0.0\\
& A9 & T17 & 33.0 &  9.9 &  73.5 & 0.0\\
\end{tabular}
\end{ruledtabular}
\end{table*}

To investigate the effect of the structural fluctuations on the
calculated couplings between the adenine and thymine basepairs we
analyzed 4000 conformations of selected basepairs from the 80 ps
molecular dynamics trajectories of the $(A)_{12}\cdot(T)_{12}$. 
The extracted 4000 snapshots span the whole 80 ps simulations with 
each snapshot taken every 20 fs. In Table ~\ref{table3}) 
the average values of the couplings calculated using IDA and TDC 
methods for selected pairs of the $(A)_{12}\cdot(T)_{12}$ oligomer 
are listed. The comparison of these values with corresponding 
couplings calculated for basepairs in their idealized B-DNA geometry 
(Table~\ref{table2}) shows some very interesting points. 
Comparing the maximum and minimum values of the coupling elements in 
Table~\ref{table3} it can be seen that the magnitude of the couplings 
is very sensitive to the structural fluctuations observed in the MD simulations. 
The absolute value of the coupling can differ by as much as 1000 cm$^{-1}$ 
for ${\pi}$-stacked nucleobases. Other arrangements of the bases exhibit 
much smaller fluctuations of the couplings in the range of 300 cm$^{-1}$ 
for Watson-Crick basepairs and still less for cross terms between bases 
on opposite strands and second nearest neighbors located on the same 
strand (Table~\ref{table3}). The fluctuations of the couplings observed for 
the intrastrand nearest neighbors show also more complex pattern compared to 
these for Watson-Crick basepairs. An example is given by the couplings
calculated with TDC method for the A9-A10 step (data not shown) of
$(A)_{12}\cdot(T)_{12}$ oligomer. For this base pair
the relatively larger and slower fluctuations on appoximately 20 ps
time scale are superimposed on the rapid fluctuations. Similar slower
fluctuations are observed for other ${\pi}$-stacked adenines but not
Watson-Crick pairs. For both ${\pi}$-stacked adenines and thymines the
Coulombic coupling seems to also depend on the location of a given pair
along the chain. As can be inferred from data in (Table~\ref{table3}) 
the base pairs closer to the 3' end, A9-A10 and T21-T22 show larger
average values of the couplings compared with the corresponding values
calculated for the base pairs closer to the 5' end. 

In spite of the large fluctuations of the couplings the values averaged
over 4000 conformations are in good agreement with the corresponding
couplings calculated for basepairs in standard B-DNA geometry. The
best agreement is observed for the Watson-Crick base pairs while somewhat 
worse is seen for the ${\pi}$-stacked thymines. Interestingly, the average 
magnitudes of couplings for diagonal base pairs in the 5' and 3' directions 
are very similar when calculated using IDA method. The average couplings 
of the A8-T16 and A9-T17 base pairs calculated with this method for the 4000 
conformations from the MD simulations are equal to 129 and 122 cm$^{-1}$, 
respectively (Table~\ref{table3}). The transition density cubes calculated 
couplings, however still show sensitivity the the direction with the average 
coupling in the 5' direction for A8-T16 , 77 cm$^{-1}$, compared to only 
33 cm$^{-1}$ in 3' direction for A9-T17.

\subsection{Exciton states}
In this section we compare the properties of the excited states of the
double-stranded DNA model (A)$_{12}\cdot$(T)$_{12}$
calculated in the framework of the exciton theory where the
off-diagonal elements of the exciton matrix were calculated using the
ideal dipole approximation (IDA) and transition density cubes (TDC).
The transition energies, oscillator strength, and the localization of
the excited states were determined by diagonalization of the  exciton
matrix with the transition energies of the individual bases (diagonal
terms) obtained from TDDFT calculations using standard geometries of 
the nucleobases while the coupling elements
(off-diagonal terms) were determined using either IDA or TDC method.
All properties were averaged over an ensamble of 240 conformations
extracted form the MD simulations.

\begin{figure}[t]
\includegraphics[width=\columnwidth]{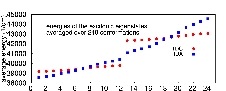}
\caption{Eigenstate energy of (A)$_{12}$(T)$_{12}$ averaged over 240 conformations from MD simulations obtained
using TDC (filled circles) and IDA (filled squares).}
\label{fig10}
\end{figure}

Fig.~\ref{fig10} shows the average values of the energies of the 24 eigenstates
of the $(A)_{12}\cdot(T)_{12}$ obtained with the
couplings calculated using TDC and IDA method. Using the former method
for calculations of the coupling elements the energies of the lowest 12
eigenstates are clearly separated from those of the remaining twelve
eigenstates. The energy change between the two border eigenstates, 
{\textless}12{\textgreater} and {\textless}13{\textgreater},
amounts to almost 2500 cm$^{-1}$ while the difference between the highest
and the lowest energy eigenstates in each of the two sets is less than 1000
cm$^{-1}$. The variations in the energy of a given
eigenstate do not exceed 150 cm$^{-1}$. The abrupt
energy change between the two sets of eigenstates diminishes
significantly when the IDA method is employed for calculations of the
dipolar couplings (Fig.~\ref{fig10}). The difference between the average
energies of the border eigenstates {\textless}12{\textgreater} and 
{\textless}13{\textgreater} is less than 700 cm$^{-1}$ with variations 
in the energies of these two eigenstates equal to 420 and 350 cm$^{-1}$, 
respectively. 

\begin{figure}[t]
\includegraphics[width=\columnwidth]{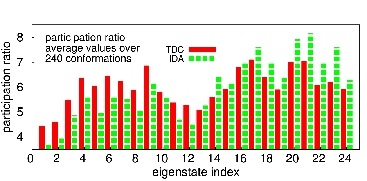}
\caption{Participation ratio of the 24 eigenstates of the 
  (A)$_{12}$(T)$_{12}$. The values averaged over 240
  conformations from MD simulations were calculated using TDC 
  (solid lines) and IDA(dashed lines) methods.}
\label{fig11}
\end{figure}

\begin{figure}[h]
\includegraphics[width=\columnwidth]{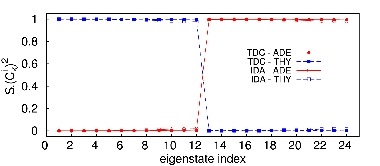}
\caption{Contribution of the $S_{2}$ states of thymine and adenine 
  to the eigenstates of (A)$_{12}$(T)$_{12}$ oligomer.}
\label{s2_contrib}
\end{figure}

The spatial extent of the eigenstates was evaluated based on the
inverse participation ratio (${IPR}$) of a given eigenstate (Eq.~\ref{ipr}),
which indicates the number of coherently bound chromophores
\cite{Bouvier2002vu}. The plot in Fig.~\ref{fig11} shows the average ${IPR}$ 
values for each of the 24 eigenstates calculated using TDC and IDA to 
obtain the couplingelements of the Hamiltonian matrix. The average values 
of the inverse participation ratios for (A)$_{12}\cdot$(T)$_{12}$ obtained 
with TDC are in the range between 4.5 and 7.1. The corresponding values 
obtained using TDC method are slightly smaller for the first 12 eigenstates 
and slightly larger for the remaining 12 eigenstates and are in the range 
between 3.7 and 8.2. In both cases the ${IPR}$ values are much larger than 
one indicating delocalization of the excitation over several bases. 
Markovitsi {\em et al.} \cite{Emanuele:2005,Markovitsi:2006,Emanuele:2005a} 
showed that for the columnar aggregates of $n$ identical chromophores, 
the maximum values of the normalized inverse participation ratio $1/{nL_{k}}$ 
is equal to 0.7. Therefore, in case of the $(A)_{12}\cdot(T)_{12}$ oligomer 
a completely delocalized eigenstate over one strand of the
double helix would have a participation ratio equal to 8.4. Contrary
to what Bouvier reported for the (A)$_{20}\cdot$(T)$_{20}$ and 
(dAdT)$_{10}\cdot$(dAdT)$_{10}$ oligomers, we found the participation ratios 
for all the eigenstates to be lower than 8.4. Therefore, a delocalization of the
eigenstates over only adenosine or thymine chromophores but not both is
expected. As can be seen from the contribution of the $S_{2}$ transitions 
of adenine and thymine to the eigenstates of the (A)$_{12}\cdot$(T)$_{12}$ 
(Fig.~\ref{s2_contrib}) the lower energy eigenstates are localized almost 
completely on the $S_{2}$ transition on the thymine, while the higher energy 
eigenstates are localized on the $S_{2}$ transition of adenosine.

\begin{figure}[t]
\centering
\includegraphics[width=\columnwidth]{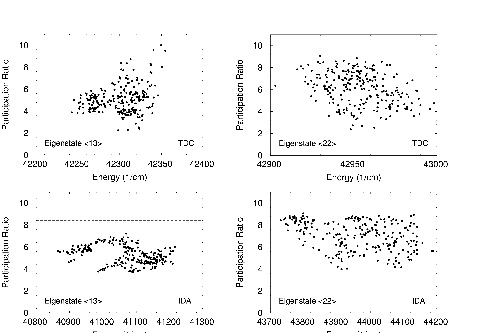}
\caption{Plot of the participation ratio of the eigenstate numbers 
        {\textless}13{\textgreater} and {\textless}22{\textgreater} 
         as a function of energy determined for 240 conformations of 
         (A)$_{12}$(T)$_{12}$.}
\label{PRvsE}
\end{figure}

The inverse participation ratios of the eigenstates {\textless}13{\textgreater} 
and {\textless}22{\textgreater} as a funcion of energy are ploted in Figure 
\ref{PRvsE}. The {\em IPR} values for these two eigenstates of (A)$_{12}$(T)$_{12}$ 
calculated for 240 conformations taken from the MD simulations show large 
fluctuactions in the range of $2-10$. Despite the wide range of calculated {\em IPR}
values, as can be seen from the plots in Figure \ref{PRvsE}, the higher 
energy eigenstate, {\textless}22{\textgreater}, on average shows larger 
delocalization compared with the lower energy eigenstate, {\textless}13{\textgreater},
wheather the IDA or TDC method was used to calculate coupling elements. 
However, only for a handful conformations the value of {\em IPR} exceeds the 
theoretical value of 8.4 (indicated by a dashed line in Figure \ref{PRvsE}) 
corresponding to the completely delocalized exciton over one strand of the 
(A)$_{12}$(T)$_{12}$. This indicates that both eigenstates {\textless}13{\textgreater} 
and {\textless}22{\textgreater} which are localized on the transition associated 
with adenine remains localized on only one strand of the (A)$_{12}$(T)$_{12}$ 
composed of adenine nucleobases.

\begin{figure}[t]
\centering
\includegraphics[width=\columnwidth]{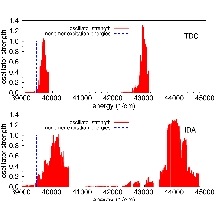}
\caption{Oscillator strength calculated for 24 eigenstates of the (A)$_{12}$(T)$_{12}$.
          The off-diagonal elements of the exciton matrix were calculated using the TDC
          (top) and IDA (bottom) methods. The values were averaged over 240 conformations
           extracted from the MD simulations.}
\label{fosc}
\end{figure}

The average values of the oscillator strengths versus the average energies 
of the eigenstates for 240 conformations from MD simulation are plotted in 
Figure ~\ref{fosc}. The total oscillator strength is distributed over a 
small number of eigenstates clustered in two bands. The first one comprise 
eigenstates {\textless}9{\textgreater} to {\textless}12{\textgreater} 
localized on thymine strand, and the second eigenstates {\textless}21{\textgreater} 
to {\textless}24{\textgreater} on adenine strand. The corresponding energies 
are around 39700 and 43000 cm\textsuperscript{ -1} for the off-diagonal 
couplings obtained using TDC and 40100 and 44000 cm\textsuperscript{-1} 
for off-diagonal terms calculated with IDA. These ``bright'' states correspond 
to the higher energy eigenstates built on the thymine and adenosine 
monomer transitions (Figure ~\ref{fosc}), while their ``dark'' counterparts, 
carrying negligible oscillator strength, correspond to eigenstates with lower energies.
The largest oscillator strenthts are carried by the eigenstates {\textless}10{\textgreater} 
and {\textless}22{\textgreater} and in both cases (IDA and TDA) the energies of 
these eigenstates are blue shifted with respect to transition energies of the 
individual bases. The magnitude of the shift approximately 1000 cm\textsuperscript{ -1} 
and less than 500 cm\textsuperscript{ -1} for IDA and TDC, respectively, reflects 
the differences in the couplings obtained with these two methods.

\section{Conlusions and Summary}
We have investigated the properties of the excited states of
$(A)_{12}(T)_{12}$ double helix calculated in
the framework of the exciton theory. In our approach we combined the
quantum mechanical calculations with the molecular dynamics
simulations. The TDDFT calculations were employed to calculate the
energies of the singlet excited states of the individual nucleobases.
The transition moments and densities of the $S_{0}\rightarrow S_{2}$ 
transitions of adenine and thymine
which correspond to the lowest energy $\pi\pi^{*}$ transitions for
these two bases were used to calculate the off-diagonal elements of
the exciton matrix. The effect of the conformational changes were
incorporated by averaging the calculated spectral properties of the
double -stranded $(A)_{12}(T)_{12}$  over large
number of the conformations extracted from the molecular dynamics
simulations. 

The Coulombic couplings calculated using the IDA and TDC methods show a
large deviations for the distances between chromophores typical for the
DNA double helices. The magnitude of the couplings calculated with IDA
being always larger than the corresponding values obtained with TDC.
The agreement between the two methods is satisfactory only for the 
separations between the chromophores larger than 5 {\AA}. The largest 
difference between these two methods is observed for the ${\pi}$-stacked 
adenines in the standard B-DNA geometry for which the coupling calculated 
with IDA is over five times larger than the corresponding values calculated 
using TDC. The effect of the structural fluctuations on the calculated 
coupling elements is also very significant for both methods the values 
of the calculated coupling can change by an order of magnitude for different 
conformations of a given basepair. The difference between the smallest and 
largest coupling between the stacked adenines calculated using IDA for a 
given base pair can be as large as 1000 cm$^{-1}$, smaller but still 
significant difference in the range of 300 cm$^{-1}$ was calculated using TDC.

The properties of the excited states of the $(A)_{12}(T)_{12}$ calculated 
in the framework of the exciton theory are affected to a different extent 
when the off-diagonal elements of the exciton matrix calculated using IDA 
and TDC methods. The eigenstates which carry the largest the oscillator strength, 
{\textless}10{\textgreater} and {\textless}22{\textgreater}, are slightly 
blue-shifted with respect to the transition energies of single nucleobases 
(Figure \ref{fosc}). The larger shift, approximately 1000 cm\textsuperscript{-1}, 
is observed for the exciton states obtained with the off-diagonal elements of 
the exciton matrix calculated using the IDA approximation, compared to less 
than 500 cm\textsuperscript{-1} obtained with the dipolar coupling calculated 
using TDC method. However, the delocalization properties of these eigenstates 
is similar for both IDA and TDC couplings. The {\em IPR} values of the ``bright'' 
eigenstate {\textless}10{\textgreater} calculated with both IDA and TDC couplings 
are 5.5 and 6.0, respectively, while the corresponding {\em IPR} values of eigenstate {\textless}22{\textgreater} equal to 7.1 and 6.1 indicate only slightly large 
delocalization. Accordingly, comparing the {\em IPRs} obtained with TDC couplings 
the initial population of the bright eigenstates {\textless}10{\textgreater} 
and {\textless}22{\textgreater} by UV absorption will create exciton states which 
are delocalized over roughly 6 thymine and adenine bases, respectively. 
Upon relaxation the exciton states become more localized (Figures \ref{PRvsE} 
and \ref{fig11}) as indicated by lower {\em IPR} values of the border eigenstates 
{\textless}1{\textgreater} and {\textless}13{\textgreater}.

% If in two-column mode, this environment will change to single-column
% format so that long equations can be displayed. Use
%\begin{widetext}
%\end{widetext}

% figures should be put into the text as floats.
% Use the graphics or graphicx packages (distributed with LaTeX2e).
% See the LaTeX Graphics Companion by Michel Goosens, Sebastian Rahtz,
% Fill in the caption in the braces of the \caption  command. Put the label
% that you will use with \ref  command in the braces of the \label  command.

% tables follow here or maybe be put in the text
% Fill in the caption in the braces of the \caption  command. Put the label
% that you will use with \ref  command in the braces of the \label  command.
% \begin{tabular}  command.
% The ruledtabular enviroment adds doubled rules to table and sets a
% nice set of default table settings.
% Use the table* environment to get a full-width table in two-column

\begin{acknowledgments}
This work was funded in part by grants from the National Science Foundation 
and the Robert Welch Foundation.  We are also grateful to 
Dr. Gillian C. Lynch and Prof. B. Montgomery Pettitt for providing 
us with the MD simulation data. We thank Dr. Stephen Bradforth for stimulating
discussion motivating this work.
\end{acknowledgments}

\end{document}